\documentclass[aps,pra,showpacs,english,10pt,a4,nofootinbib,notitlepage,twocolumn,superscriptaddress]{revtex4-1}

\usepackage[utf8]{inputenc}
\usepackage[T1]{fontenc}
\usepackage[english]{babel}  

\usepackage{times}
\usepackage{dsfont}

\usepackage{amssymb,amsmath,amsfonts,amsthm}
\usepackage{mathtools}
\usepackage{verbatim}
\usepackage{enumerate}
\usepackage{graphicx}
\graphicspath{{figs/}}
\usepackage{hyperref}
\usepackage{bbm}
\usepackage{booktabs}

\usepackage[caption=false]{subfig}

\usepackage{color}
\definecolor{dred}{rgb}{.8,0.2,.2}
\definecolor{ddred}{rgb}{.8,0.5,.5}
\definecolor{dblue}{rgb}{.2,0.2,.8}
\definecolor{dgreen}{rgb}{.2,0.5,.2}

\newcommand{\tr}{\textrm{tr}}

\newcommand{\be}{\begin{equation}}
\newcommand{\ee}{\end{equation}}

\newcommand{\bea}{\begin{eqnarray}}
\newcommand{\eea}{\end{eqnarray}}

\usepackage{amsmath,amsthm,amsfonts,amssymb}
\usepackage{dsfont,bbm}
\usepackage{color,soul}
\usepackage{tikz}

\newcommand{\ket}{\rangle}
\newcommand{\bra}{\langle}

\usepackage[normalem]{ulem}

\begin{document}

\title{Dynamical-Invariant-based Holonomic Quantum Gates: Theory and Experiment}

\author{Yingcheng Li}
\affiliation{State Key Laboratory of Surface Physics,Department of Physics, Center for Field Theory and Particle Physics, and Institute for Nanoelectronic devices and Quantum computing, Fudan University, Shanghai 200433, China}

\author{Tao Xin}
\email{xint@sustech.edu.cn}
\affiliation{Shenzhen Institute for Quantum Science and Engineering and Department of Physics, Southern University of Science and Technology, Shenzhen 518055, China}
 \affiliation{Guangdong Provincial Key Laboratory of Quantum Science and Engineering, Shenzhen 518055, Guangdong, China}

\author{Chudan Qiu}
\affiliation{Shenzhen Institute for Quantum Science and Engineering and Department of Physics, Southern University of Science and Technology, Shenzhen 518055, China}

\author{Keren Li}
\affiliation{Center for Quantum Computing, Peng Cheng Laboratory, Shenzhen 518055, China}

\author{Gangqin Liu}
\affiliation{Institute of Physics, Chinese Academy of Sciences, Beijing 100190, China}

\author{Jun Li}
\affiliation{Shenzhen Institute for Quantum Science and Engineering and Department of Physics, Southern University of Science and Technology, Shenzhen 518055, China}
 \affiliation{Guangdong Provincial Key Laboratory of Quantum Science and Engineering, Shenzhen 518055, Guangdong, China}

\author{Yidun Wan}
\email{ydwan@fudan.edu.cn}
\affiliation{State Key Laboratory of Surface Physics,Department of Physics, Center for Field Theory and Particle Physics, and Institute for Nanoelectronic devices and Quantum computing, Fudan University, Shanghai 200433, China}
\affiliation{Shenzhen Institute for Quantum Science and Engineering and Department of Physics, Southern University of Science and Technology, Shenzhen 518055, China}

\author{Dawei Lu}
\email{ludw@sustech.edu.cn}
\affiliation{Shenzhen Institute for Quantum Science and Engineering and Department of Physics, Southern University of Science and Technology, Shenzhen 518055, China}
 \affiliation{Guangdong Provincial Key Laboratory of Quantum Science and Engineering, Shenzhen 518055, Guangdong, China}

\begin{abstract}
Among existing approaches to holonomic quantum computing, the adiabatic holonomic quantum gates (HQGs) suffer errors due to decoherence, while the non-adiabatic HQGs either require additional Hilbert spaces or are difficult to scale. Here, we report a systematic, scalable approach based on dynamical invariants to realize HQGs without using additional Hilbert spaces. While presenting the theoretical framework of our approach, we design and experimentally evaluate single-qubit and two-qubits HQGs for the nuclear magnetic resonance system. The single-qubit gates acquire average fidelity 0.9972 by randomized benchmarking, and the controlled-NOT gate acquires fidelity 0.9782 by quantum process tomography. Our approach is also platform-independent, and thus may open a way to large-scale holonomic quantum computation. 
\end{abstract}

\maketitle

\emph{Introduction.} -- In holonomic quantum computing (HQC), one controls the quantum evolution of a qubit system such that the geometric phases accumulated in the evolution passages realize a universal set of HQGs over the computational space \cite{Zanardi2012}, which are believed to be more robust against certain types of errors than usual dynamical gates \cite{Solinas2003a,Jing2017,Zhu2005,Zhou2017,Berger2013,Carollo2003}. In the original proposal of HQC, HQGs are adiabatic and have been experimentally implemented in nuclear magnetic resonance (NMR) \cite{Jones2000a} and superconducting circuits \cite{Falci2000}. Unfortunately, adiabatic HQGs operate too slowly to ignore  decoherence. To speed up HQGs, non-adiabatic HQGs were proposed \cite{Aharonov1987a,Berry2009a,Xu2012,Sjoqvist2012} by adding ancillary Hilbert space in addition to the computational space\cite{Feng2013a,Abdumalikov2013,Zu2014,Wendin2017a,Xu2018a,Yan2019,Liu2019b,Danilin2018,Li2020,Qi2020}. In the original realizations of HQGs, control passages confined over the $n$-qubit computational subspace form a discrete set, leading to the difficulty in locating the easily implementable control passages. Because of the ancillary Hilbert space, the total Hilbert space bears infinite control passages that form a continuous hypersurface. Hence, it is much easier to find a control passage to realize the HQGs in the computational subspace; however, the cost is that the additional Hilbert space will cause leakage and sometimes lengthen the HQGs. Proposals that do not use ancillary Hilbert spaces\cite{Chen2018a,Kleibler2018,Xu2019} usually have to meticulously design the shape of an evolution passage in the Hilbert space. Such a design is difficult to achieve in  Hibert spaces higher than three-dimensional. Therefore, to reduce the complexity  and error sources in its physical realization, HQC begs a systematic method of implementing any-qubit HQGs without using an ancillary Hilbert space. 

In this work, we develop a systematic approach to multi-qubit non-adiabatic HQGs without an ancillary Hilbert space, by means of the dynamical-invariant-based quantum control. As an application of our approach, we design the HQGs for NMR systems and experimentally test them in our NMR system. The notion of dynamical invariants (DIs) was proposed by Lewis and Reisenfeld in 1969 to solve the time-dependent Schr{\"o}dinger equation analytically, such that any solution to the Schr\"odinger equation is a superposition of the instantaneous eigenstates of the DI of the Hamiltonian \cite{Lewis1969a}. If a quantum system is driven to evolve in certain instantaneous eigenstates of its dynamical invariant, the control is non-adiabatic. About a decade ago, Chen \it{et al} \rm proposed a non-adiabatic quantum control method, called inverse engineering, for two-level systems based on DIs \cite{Chen2010,Chen2011a,Chen2010a}, but the method was difficult to scale up beyond two-level systems \cite{Fasihi2012b}. Later, Gungordu \it et al \rm classified the DIs of generic $N$-level systems using a Lie-algebraic method \cite{Gungordu2012b} and proposed DI-based HQGs \cite{Gungordu2014a}; however, Ref. \cite{Gungordu2014a} could not offer a systematic method of designing other $n$-qubit HQGs because the DI equation for a generic $n$-qubit Hamiltonian is difficult to solve analytically.

We show that under reasonable assumptions, the differential equations of the DIs of a system can be converted into linear equations, enabling us to write down the closed-form DI-based unitary evolution operator of that system. Based on the closed-form evolution operator, we develop a systematic approach that turns the problem of designing HQGs into a program-solvable problem of maximizing what we call a fidelity function. As an example, we demonstrate that our method is effective for NMR-type Hamiltonians, which comprise single-qubit radio-frequency (RF) pulse terms, single-qubit Zeeman terms, and Ising-type coupling terms. We design and experimentally implement the non-adiabatic holonomic single-qubit gates (including the NOT, Hadamard, phase, and $\frac{\pi}{8}$ gates) and the two-qubit CNOT gate without any ancillary qubits in an NMR quantum processor. Our single-qubit gates are implemented with fewer pulses than before (\cite{Gungordu2014a}) and result in fidelity with all gates over $99\%$. On top of that, the CNOT gate achieves fidelity $97.8\%$. Our method of designing non-adiabatic HQGs is also platform-independent, i.e., applicable to other quantum systems, such as the defects in diamond and superconducting circuits. We shall report the results on such systems elsewhere.

\emph{DI-based HQC.} -- We first introduce the DIs and how they lead to non-adiabatic HQGs. For a time-dependent Hamiltonian $\mathcal{H}(t)$, a corresponding dynamical invariant $\mathcal{I}(t)$ is a time-dependent Hermitian operator with constant expectation value and thus satisfies the following DI equation\cite{Lewis1969a}
\begin{equation}\label{DI}
\frac{\partial\mathcal{I}(t)}{\partial t}+i\left[\mathcal{H}(t),\mathcal{I}(t)\right]=0.
\end{equation}
As shown in ref. (\cite{Lewis1969a}), an analytic solution $|\psi(t)\ket$ to the time-dependent Schr{\"o}dinger equation can be expanded by the instantaneous eigenstates of the DI $\mathcal{I}(t)$,
\begin{equation}\label{Sol}
|\psi(t)\ket=\sum_n c_n e^{i\alpha_n(t)}|\varphi_n(t)\ket,
\end{equation}
where $c_n$'s are time-independent complex constants, and
\begin{equation}\label{Phase}
\alpha_n(t)=\int_0^t \bra\varphi_n(s)| i\frac{\partial}{\partial s}-H|\varphi_n(s)\ket ds.
\end{equation}

In terms of the instantaneous eigenstates of the DI, the unitary evolution operator of the Hamiltonian $\mathcal{H}(t)$ is written as
\begin{equation}\label{Unitary}
U(t)=\sum_n e^{i\alpha_n(t)}|\varphi_n(t)\ket \bra \varphi_n(0)|.
\end{equation}
Since an eigenstate of $\mathcal{I}(0)$ evolves in the form
\begin{align}
U(t) |\varphi_n(0)\ket=e^{i\alpha_n(t)}|\varphi_n(t)\ket,
\end{align}
which is transitionless in the eigenbasis $\{|\varphi_n(t)\ket\}$ but transits among eigenstates of $\mathcal{H}(t)$, and hence is not limited by the adiabaticity condition. The unitary evolution operator is an exact solution to the Schr{\"o}dinger equation due to Eq. (\ref{Sol}).

In a cyclic evolution, the phase factor in Eq. (\ref{Phase}) can be separated into two parts. The geometric phase (or the holonomy)
\begin{align}\label{Geophase}
\gamma^g_n=\int_0^T i \bra \phi_n(s) | \frac{d}{ds}|\phi_n(s)\ket ds=\oint i \bra \phi(t) | d | \phi(t) \ket
\end{align}
is determined by the Berry connection in the Hilbert space and depends on the trajectory of the instantaneous eigenstates of the DI in the Hilbert space.
The dynamical phase
\begin{align}\label{Dphase}
\gamma^d_n=-\int^T_0\bra \phi_n(s)| H(s) |\phi_n(s)\ket ds,
\end{align}
on the other hand, is sensitive to the evolution `velocity' of the instantaneous eigenstates of the DI in the Hilbert space. When the dynamical phase vanishes, viz $\gamma^d_n=0$, the geometric phase $\gamma^g_n$ fully determines the cyclic evolution operator, which is nontrivial and in fact an HQG.

\emph{General framework.} -- Here we show how to employ the DI-based approach to design non-adiabatic HQGs efficiently and platform-independently. We treat the closed-form formula of the evolution operator \eqref{Unitary} and the dynamical phase \eqref{Dphase} as functions of the parameters in the Hamiltonian. Then, we search for the values of these parameters that not only annihilate the dynamical phase but also render the evolution operator being certain quantum gates. Therefore, we need to solve the DI equation to derive the closed-form formula of the evolution operator and the dynamical phase for the systems of concern. We start from the general Hamiltonian that is commonly used in quantum computing systems,
\begin{align}\label{nbitHam}\notag
\mathcal{H}_n=&\frac{1}{2}\sum_{i=1}^n (\Omega_i \cos(\omega_i t+\phi_i)\sigma_x^i+\Omega_i \sin(\omega_i t+\phi_i) \sigma_y^i))
\\
+&\frac{1}{2}\sum_{i=1}^n \Delta_i \sigma_z^i+\frac{1}{4}\sum_{i<j}J_{ij}\sigma_z^i \sigma_z^j,
\end{align}
where $\Delta_i$ is the strength of the Zeeman energy, and $\Omega_i$ and $\omega_i$ are the amplitude and frequency of the control field for each qubit, respectively. The generating set of this Hamiltonian is $\bigcup_{i=1}^n\{\sigma_x^{i}, \sigma_y^{i}, \sigma_z^{i}\}
\cup\bigcup_{i<j}\{\sigma_z^{i}\sigma_z^{j}\}$, which spans the entire $su(2^n)$ Lie algebra. In other words, the corresponding DIs in general have $4^n-1$ terms, which give rise to $4^n-1$ differential equations that cannot be solved by separating variables. Nevertheless, for our purpose of building non-adiabatic HQGs, we only need one special solution to these equations. Since the Hamiltonian in Eq. (\ref{nbitHam}) has only $\frac{n(n+5)}{2}$ terms, the $4^n-1$ terms in the DI are redundant and some of them can be assumed zero. In the DI, if the Cartan sub-algebra generators, viz $\sigma_z^{i}$'s and $\sigma_z^{i}\sigma_z^{j}$'s, have time-independent coefficients, these Cartan terms will not survive the time-derivative in the DI equation. The mixing terms that contain $\sigma_x^{i}\sigma_z^{j}$, $\sigma_y^{i}\sigma_z^{j}$, $\sigma_z^{i}\sigma_x^{j}$, and $\sigma_z^{i}\sigma_y^{j}$ should also vanish because these terms will become the Cartan terms that cannot be cancelled after commuting with the Hamiltonian. Moreover, the time-dependent part of the DI should have sinusoidal time dependence, such that they can eliminate the sine terms in the Hamiltonian. Hence, a simple solution to the DI equation has a form similar to the Hamiltonian (\ref{2bitHam}), which consists of a Zeeman term, Ising term, and sine control-field term with the same frequency,
\begin{align}\label{nbitDI}\notag
\mathcal{I}_n=&\sum_{i=1}^n (\Omega_i \cos(\omega_i t+\phi_i)\sigma_x^i+\Omega_i \sin(\omega_i t+\phi_i) \sigma_y^i)
\\
+&\sum_{i=1}^n (\Delta_i-\omega_i) \sigma_z^i+\frac{1}{2}\sum_{i<j}J_{ij}\sigma_z^i \sigma_z^j.
\end{align}
A proof that $\mathcal{I}_n$ and $\mathcal{H}_n$ satisfy the DI equation can be found in the Supplementary Material \ref{appd:proof}.

Denote the set of parameters in the Hamiltonian by a vector $\mathbf{R}$. To realize a specific gate $U_0$, we solve the eigen-problem of the DI $\mathcal{I}_n$, and derive the closed-form formula of the evolution operator $U(\mathbf{R})$ and the dynamical phase $\gamma^d_n(\mathbf{R})$ using Eqs. (\ref{Unitary}) and (\ref{Dphase}). One may need multiple cyclic evolutions---the total evolution operator will be $\prod_i U(\mathbf{R}_i)$---to realize certain gates. We optimize the parameter set for each $U(\mathbf{R}_i)$ and maximize the fidelity function $F=\tr( U_0^\dagger\cdot  \prod_i U(\mathbf{R}_i) )$, while setting $\gamma^d_n(\mathbf{R}_i)=0$. The parameter set that maximizes the function $F$ eventually makes the total evolution operator $\prod_i U(\mathbf{R}_i)$ the desired HQG $U_0$. Note that maximizing the fidelity function permits multiple solutions, which may suit for situations requiring different gate lengths (detailed in Supplemental Information B). 
We remark that changing the basis of the Pauli matrices provides more degrees of freedom to reduce the number of cyclic evolutions required for a gate.
Besides, setting any of the parameters $\Omega_i$, $\phi_i$, and $\Delta_i$ to zero maintains the DI equation; hence, the $2$-qubit unitary evolution operator covers all the cases that are discussed in (\cite{Gungordu2012b}). 

We now demonstrate in NMR systems, as an application of our method, how to realize very fast single-qubit gates and two-qubit gates.

\emph{DI-based HQGs in NMR.} The general single-qubit Hamiltonian in NMR is
\begin{equation}\label{1bitHam}
\mathcal{H}_1=\frac{1}{2}(\Omega \cos(\omega t+\phi)\sigma_x+\Omega \sin(\omega t+\phi)\sigma_y+\Delta \sigma_z),
\end{equation}
where $\Omega$ and $\omega$ are the amplitude and frequency of the control field, respectively, and $\Delta$ is the strength of the Zeeman energy.

The corresponding DI equation for the Hamiltonian, according to Eq. (\ref{nbitDI}), is,
\begin{equation}\label{1bitDI}
\mathcal{I}_1=\Omega \cos(\omega t+\phi)\sigma_x+\Omega \sin(\omega t+\phi)\sigma_y+(\Delta-\omega) \sigma_z.
\end{equation}
The instantaneous eigenstates of the $\mathcal{I}_1$ are Bloch states with fixed precession frequency $\omega$, fixed cone angle $\theta=\arctan(\frac{\Omega}{\Delta-\omega})$, and initial phase $\phi$. 

An eigenstate acquires a geometric phase $\gamma^g=\pm\pi(1-\mathrm{cos}\frac{\theta}{2})$ after a period $T=\frac{2\pi}{\omega}$. To vanishing dynamical-phase condition $\gamma^d_n=0$ gives
\begin{equation}\label{condition}
\Omega^2+\Delta(\Delta-\omega)=0.
\end{equation}
 Constrained by condition (\ref{condition}), the unitary evolution of the system Hamiltonian becomes
\begin{align}\label{1bitgate}
U^g(\theta,\phi)=-e^{-i\pi\cos\theta(\sin\theta\cos\phi\sigma_x+\sin\theta\cos\phi\sigma_y+\cos\theta\sigma_z)}.
\end{align}
In experiment, the magnetic field needs to `jump' between two cyclic evolutions. These `jumps' cause the dominating errors, and hence the more number of cycles are involved, the more errors are caused. To minimize the `jump' errors between different cyclic evolutions, we search the values of the parameters until the smallest number of cyclic evolutions is found. It turns out that any single-qubit gate can be realized within two cyclic evolutions. The corresponding parameters for the NOT, Hadamard, phase, and $\pi/8$ gates are listed in the Supplementary Information \ref{appd:value}. The parameters listed in Supplementary Information \ref{appd:value}, as aforementioned, are not unique.

Having realized the necessary holonomic single-qubit gates in NMR, we now move on to  two-qubit gates, whose Hamiltonian contains the system Hamiltonian and the control field:
\begin{align}\label{2bitHam}
\mathcal{H}_2&= \sum_{i=1}^2 \Omega_i (\mathrm{cos}(\omega_i t+\phi_i)\frac{\sigma_x^{i}}{2}+ \mathrm{sin}(\omega_i t+\phi_i)\frac{\sigma_y^{i}}{2} )  \\ \nonumber
& + \Delta_1 \frac{\sigma_z^{1}}{2} +\Delta_2 \frac{\sigma_z^{2}}{2}   + J\frac{\sigma_z^{1}\sigma_z^{2}}{4} ,
\end{align}
The solution to the DI equation is
\begin{align}\label{2bitDI}
\mathcal{I}_2&= \sum_{i=1}^2 \Omega_i (\mathrm{cos}(\omega_i t+\phi_i)\sigma_x^{i}+ \mathrm{sin}(\omega_i t+\phi_i)\sigma_y^{i} )  \\ \nonumber
& + \sum_{i=1}^2 (\Delta_i- \omega_i ) \sigma_z^{i} + J\sigma_z^{1}\sigma_z^{2}.
\end{align}
The condition of cancelling the dynamical phase contains the analytic solution to a quartic function, which is tedious and unimportant to the main story. Note that defining a cyclic evolution requires $m \omega_1=n\omega_2$, where $(m,n)$ is a pair of integers. For simplicity, we will choose $(m,n)=(1,1)$ hereafter, such that $\omega_1=\omega_2=\omega$, but keep in mind that we still have a rational-number freedom that enables us to further reduce the number of cyclic evolutions required to realize certain gates.

Before we design the CNOT gate, we show that our method can efficiently find entangling gates with single loop. We denote the second smallest singular value of the matrix $C_{ij}=\mathrm{tr}(U_2 \sigma_i \otimes \sigma_j)$ as $M$, where $U_2$ is the evolution operator of $\mathcal{H}_2$, and $\sigma_i = I,\sigma_x,\sigma_y, \sigma_z$ for $i=1,2,3$ and $4$. The $M$ is a function of the parameters in the Hamiltonian $\mathcal{H}_2$ has a lower bound zero. Hence, minimizing the function $M(\Omega_1,\Omega_2,w_1,\phi_1,\phi_2,\Delta_1,\Delta_2)$ will lead to an evolution operator whose $C_{ij}$ matrix is rank two, which means that the evolution operator is an entangling gate. Below we show one set of parameters that realizes a generic entangling gate.

\begin{table}[!h]
\begin{tabular}{c|ccccccc}
\hline
Pulse & $\Omega_1/J$  & $\Omega_2/J$ & $\omega/J$ & $\phi_1$ & $\phi_2$ & $\Delta_1/J$ & $\Delta_2/J$ \\
\hline
P1 &0.0000&2.7610&15.0000&5.7264&0.0000&0.5000&0.5002\\
\hline
\end{tabular}
\caption{\footnotesize{Parameters in the single cyclic evolution to realize an DI-based entangling gate.}}
\end{table}

For the CNOT gate, instead of imposing the condition of cancelling the dynamical phase, we write down the two-qubit evolution operator and maximize the fidelity function while setting $\gamma^d_n(\mathbf{R}_i)=0$. The optimization process is similar to the single-qubit case. The result shows that five cyclic evolutions are sufficient for the CNOT gate, and the relevant parameters are listed in Table. \ref{CNOTpara}.

\begin{table}[!h]
\begin{tabular}{c|ccccccc}
\hline
Pulse & $\Omega_1/J$  & $\Omega_2/J$ & $\omega/J$ & $\phi_1$ & $\phi_2$ & $\Delta_1/J$ & $\Delta_2/J$ \\
\hline
P1 &1.446&4.131&8.478&3.111&1.590&0.268&4.168\\
\hline
P2 &1.956&3.819&7.837&4.437&1.431&0.561&3.761\\
\hline
P3 &3.394&4.339&8.745&2.053&3.467&1.836&3.702\\
\hline
P4 &1.807&3.591&7.394&5.127&4.532&0.510&3.555\\
\hline
P5 &2.551&4.015&8.183&1.172&4.864&0.967&3.797\\
\hline
\end{tabular}
\caption{\footnotesize{Parameters in the five cyclic evolutions to realize the DI-based CNOT gate, which minimizes the "jump" errors between two evolutions. }} \label{CNOTpara}
\end{table}

\begin{figure}[htb]
\begin{center}
\includegraphics[width=\columnwidth]{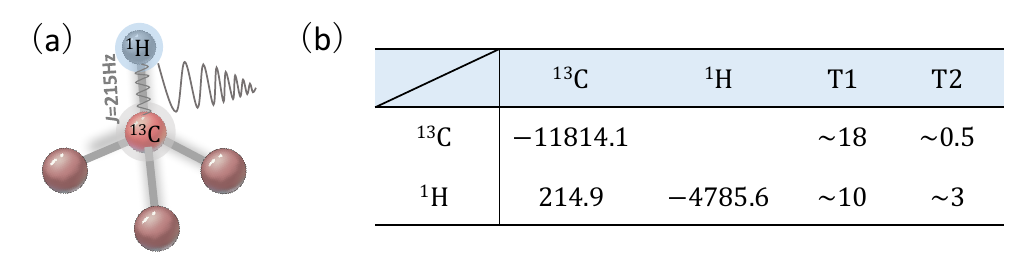}
\end{center}
\setlength{\abovecaptionskip}{-0.00cm}
\caption{\footnotesize{(a) Molecular structure and (b) parameters of the $^{13}$C-labeled chloroform. Diagonal elements and off-diagonal elements list the chemical shifts (Hz) and coupling strength (Hz) between the two spins of the molecule, respectively. The relaxation time $T_1$ and $T_2$ in the unit of seconds are determined by the standard inversion recovery and Hahn echo sequences. }}\label{mole}
\end{figure}

\emph{Experiment.} -- We implement the DI-based HQGs using the $^{13}$C-labeled chloroform sample, which servers as a 2-qubit NMR quantum processor. The nuclear spins $^{13}$C and $^{1}$H are the two qubits.  In the double-rotating frame, the internal Hamiltonian reads $\mathcal{H}_{\rm int}=\sum^2_{i=1} (\nu _i-\nu^o _i)\frac{\sigma_z^i}{2}+J\frac{\sigma_z^1 \sigma_z^2}{4}$, where $\nu _i$ and $\nu^o _i$ are the chemical shift and the reference (rotating frame) frequency of the $i$-th spin, and $J$ is the coupling strength between $^{13}$C and $^{1}$H. Compared to Eq. (\ref{2bitHam}), the required Zeeman energies $\Delta_1$ and $\Delta_2$ can be realized by varying the  detuning frequency $\mathcal{D} = \nu _i-\nu^o _i$. The molecular structure and parameters can be found in Fig. \ref{mole}.

One can control each of the two spins individually  with the RF pulse, and realize arbitrary single-qubit and two-qubit operations aided by the $J$-coupling.  The control Hamiltonian of the RF pulse reads $\mathcal{H}_{\rm c}=\sum_{i=1}^2 B_i (\mathrm{cos}(\omega_i t+\phi_i)\frac{\sigma_x^{i}}{2}+ \mathrm{sin}(\omega_i t+\phi_i)\frac{\sigma_y^{i}}{2} )$, where $B_i$, $\omega_i$, and $\phi_i$ are the amplitudes, frequencies, and phases of the RF pulse respectively. One can see that the total Hamiltonian $\mathcal{H}_{\rm int} + \mathcal{H}_{\rm c} $ with adjustable control parameters can realize the single- and two-qubit Hamiltonians in Eqs. (\ref{1bitHam}) and (\ref{2bitHam}). 

\begin{figure}[htb]
\begin{center}
\includegraphics[width=0.85 \columnwidth]{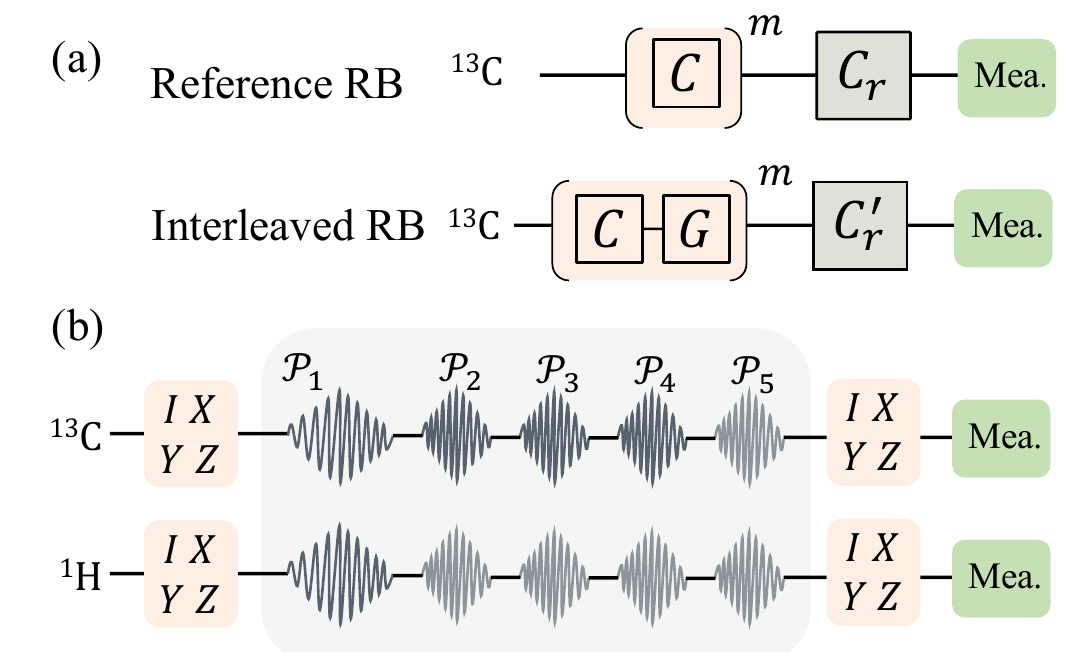}
\end{center}
\setlength{\abovecaptionskip}{-0.00cm}
\caption{\footnotesize{(a) Single-qubit RB sequence. The reference RB sequence is performed by applying $m$ random Clifford gates $C$ and a recovery gate $C_r$. The interleaved RB is performed by interleaving the target gate $G$ into the $m$ random Clifford gates. The fidelity of $G$ is calculated by $F_{G}=1-(1-p_{\rm gate}/p_{\rm ref})/2$, with the sequence decay $p_{\rm ref}$ for the reference RB and $p_{\rm gate}$ for the interleaved RB. (b) Two-qubit QPT sequence.  We  prepare the system into a 2-qubit Pauli operator, e.g. $IX$, and apply the CNOT gate (including five RF pulses labeled from $\mathcal{P}_1$ to $\mathcal{P}_5$). Quantum state tomography in the Pauli basis is performed on the final state. The matrix form of the target gate can be fully reconstructed by traversing the input from $II$ to $ZZ$. }}\label{RB}
\end{figure}

For the single-qubit DI-based HQG, we experimentally decouple the $^{13}$C from $^{1}$H and demonstrate four important single-qubit gates on the $^{13}$C, i.e., the NOT gate $X$, Hadamard gate $H$, phase gate $P$, and $\pi/8$ gate $T$. They are implemented by applying two successive RF pulses $\mathcal{P}_1$ and $\mathcal{P}_2$, where $\mathcal{P}_i$ is characterized by a set of the parameters including the detuning frequency $\mathcal{D}_i$, the control pulse $B_i$, $\omega_i$ and $\phi_i$, and the pulse duration $\tau_i$. The parameters (see Supplementary Information \ref{appd:value} for their values) are determined according to the optimization process.
We also realize the $2$-qubit CNOT  gate by concatenating five RF pulses (parameters in Table. \ref{CNOTpara}).

\emph{Results.} -- To characterize the performance of the DI-based HQGs, we implement quantum process tomography (QPT) for both single- and two-qubit gates. We also perform randomized benchmarking (RB) for single-qubit non-adiabatic holonomic (NAH) gates. Experimental sequences for QPT and RB are shown in Fig. \ref{RB}. 

\begin{figure}[htb]
\begin{center}
\includegraphics[width=1 \columnwidth]{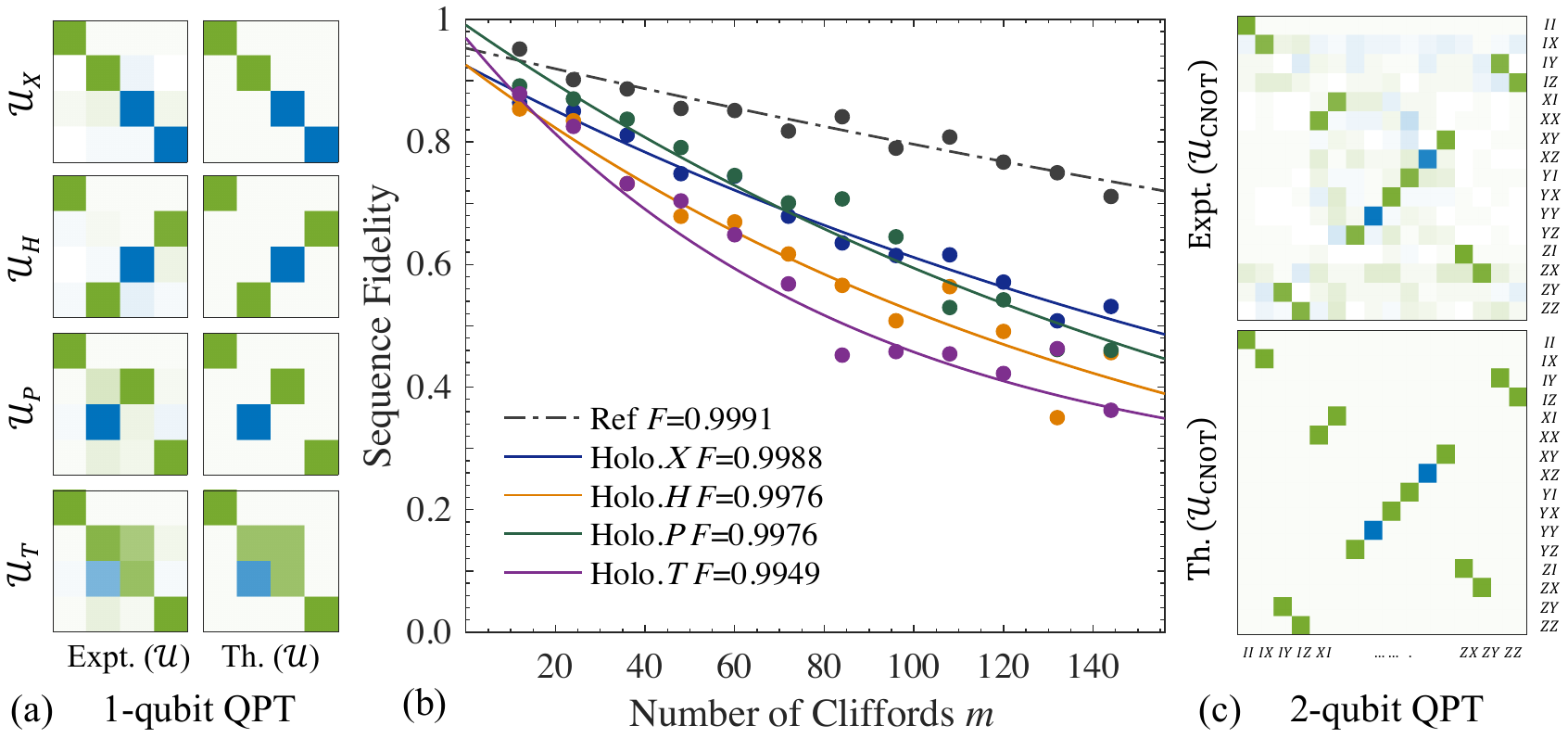}
\end{center}
\setlength{\abovecaptionskip}{-0.00cm}
\caption{\footnotesize{Experimental QPT and RB results for the single- and two-qubit DI-based HQG gates. (a) Single-qubit QPT result. The comparison between the experimental and theoretical form is given by the matrix form in the Pauli basis. (b) Single-qubit RB result. The sequence fidelity is decayed as a function of the number of Clifford gates $m$. (c) Two-qubit QPT result for the CNOT gate. The colormap ranges from -1 (the blue) to 1 (the green). }}\label{re}
\end{figure}

For single-qubit DI-based HQGs, we firstly implement traditional QPT for the four gates. The pulse lengths are $\tau_{X}=240~\mu$s, $\tau_{H}=296~\mu$s, $\tau_{P}=268~\mu$s, and $\tau_{T}=288~\mu$s. These pulses can shortened further  by increasing the detuning frequency $\mathcal{D}$. The fidelities of these four gates via QPT experiments are respectively 0.9960, 0.9953, 0.9916, and 0.9924. Note that these fidelities are usually smaller than the "pure" fidelity of the gates, as QPT cannot avoid errors in state preparation and measurement. Figure \ref{re}(a) shows the matrix forms of the four reconstructed quantum processes in the Pauli basis with comparison to the theoretical values. To test the robustness to decoherence of the DI-based HQGs, we also lengthen the gates (up to 10 ms) and perform QPT. The fidelity is at least 0.9908 for each gate even at the presence of long pulses.

RB is also performed to evaluate the performance of the single-qubit gates. In RB experiments, we initialize the system onto a fixed input state $Z$ and measure the average fidelity of the sequence after randomly repeating 40 different sequences. Figure \ref{re}(b) presents the decay of sequences with the number $m$ of Clifford gates for reference and interleaved RB sequences. Results show that the fidelity of the reference gates is $F_{\rm ref}= 0.9991$, and the average fidelity of the four target single-qubits gates around $0.9972$.

For two-qubit DI-based HQGs, we perform two-qubit QPT to characterize the CNOT gate. This gate is realized by five successive RF pulses (total length $5.584$ ms), as limited by the $J$-coupling strength $215$ Hz. The QPT experiment gives the CNOT gate a $0.9782$ fidelity. Figure \ref{re}(c) plots its matrix form in the Pauli basis to compare with the theoretical form.

\emph{Conclusion.} -- HQC is a significant candidate for fault-tolerant quantum computing. Nevertheless, HQC requires a systematic method of implementing any-qubit HQGs without using an ancillary Hilbert space. Based on dynamical invariants, we propose a systematic approach to multi-qubit non-adiabatic HQGs without using ancillary Hilbert spaces. Our approach provides fast HQGs with high fidelity. We design and experimentally demonstrate our HQGs in an NMR system. Our method is also platform-independent, while relevant results on superconducting circuits will be reported soon.

\emph{Acknowledgments.} -- This work is supported by the National Key Research and Development Program of China (Grants No. 2019YFA0308100), National Natural Science Foundation of China (Grants  No. 11875109, No. 11905099, No. 11605005, No. 11875159  and No. U1801661),  Guangdong Basic and Applied Basic Research Foundation (Grants No. 2019A1515011383), Science, Technology and Innovation Commission of Shenzhen Municipality (Grants No. ZDSYS20170303165926217, No. JCYJ20170412152620376 and JCYJ20180302174036418),  Guangdong Innovative and Entrepreneurial Research Team Program (Grant No. 2016ZT06D348), Shanghai Municipal Science and Technology Major Project (Grant No.2019SHZDZX01). YL and TX contributed equally to this work. YW and YL are grateful to Ray Laflamme for his critical, inspiring comments and suggestions, and appreciate the hospitality of the Institute for Quantum Computing at the University and Waterloo and the Perimeter Institute for Theoretical Physics, where part of this work was done. YL also thanks Mikio Nakahara and Guanru Feng for helpful discussions.

\clearpage

\begin{appendix}
\section{Calculation of the dynamical invariant}\label{appd:detail}
Starting from the definition of DI, we can go to the Heisenberg picture and see that
\begin{equation}\label{def}
\bra\psi(0)|U^\dagger (t) \mathcal{I}(t)U(t) | \psi(0)\ket=const.=\bra\psi(0)|\mathcal{I}(0)| \psi(0)\ket,
\end{equation}
which tells us that $\mathcal{I}$ is actually a constant in the Heisenberg picture. Therefore, we have
\begin{equation}
\mathcal{I}(t)|\psi(t)\ket=\mathcal{I}(t)U(t)|\psi(0)\ket=U(t)\mathcal{I}(t)|\psi(0)\ket.
\end{equation}
In other words, any eigenstate of a dynamical invariant remains in its instantaneous eigenstate of $\mathcal{I}$ under the evolution of the Hamiltonian $H$. Taking derivative of Eq. (\ref{def}) gives
\begin{equation}
\frac{d}{dt}(U^\dagger (t) \mathcal{I}(t) U(t))=0,
\end{equation}
and by substituting the Schr{\"{o}}dinger equation $i\frac{d}{dt} U(t)=H(t)U(t)$
we get the DI equation
\begin{equation} \label{DI}
\frac{\partial \mathcal{I}(t)}{\partial t}+i\left[\mathcal{H}(t),\mathcal{I}(t)\right]=0.
\end{equation}
One property that $\mathcal{I}$ follows is
\begin{equation}
i\frac{d}{dt}(\mathcal{I} |\psi(t)\ket)=H(t)(\mathcal{I}|\psi(t)\ket).
\end{equation}
Furthermore, the eigenvalue of $\mathcal{I}$ is independent of time as long as the eigenstates of $\mathcal{I}$ are complete. By taking a derivative of the following equation,
\begin{equation}
\mathcal{I} |\phi_n (t) \ket = n |\phi_n(t)\ket,
\end{equation}
where $|\phi_n (t) \ket$ is the $n$-th eigenstate of $\mathcal{I}$, we see that
\begin{equation}\label{deriv}
\frac{\partial \mathcal{I}}{\partial t}|\phi_n(t)\ket+ \mathcal{I} \frac{\partial}{\partial t} |\phi_n (t)\ket = \frac{\partial n}{\partial t} |\phi_n(t)\ket + n \frac{\partial}{\partial t}|\phi_n(t)\ket.
\end{equation}
Multiplying both sides with $\bra \phi_n(t) |$, we get
\begin{equation}
\frac{\partial n}{\partial t}= \bra \phi_n(t)| \frac{\partial \mathcal{I}}{\partial t} |\phi_n(t)\ket.
\end{equation}
Expanding Eq. (\ref{DI})
\begin{equation}
i\frac{\partial \mathcal{I}}{\partial t} |\phi_n(t)\ket +\mathcal{I} H - n H|\phi_n(t)\ket=0,
\end{equation}
and taking the inner product with $\bra \phi_{n'}(t) |$, we arrive at
\begin{equation}
\bra \phi_{n'}(t)| i\frac{\partial \mathcal{I}}{\partial t} |\phi_n(t)\ket+(n'-n)\bra \phi_{n'}(t)|H|\phi_n(t)\ket=0,
\end{equation}
which indicates that
\begin{equation}
\bra \phi_n(t)|\frac{\partial \mathcal{I}}{\partial t} |\phi_n(t)\ket=0=\frac{\partial n}{\partial t}.
\end{equation}

\section{Parameters for single- and two-qubit gates}\label{appd:value}
For single-qubit gates, we maximize the fidelity function $F=\tr( U_0^\dagger\cdot  \prod_i U(\theta_i,\phi_i) )$ with
\begin{equation}
U(\theta,\phi)=U^g(\theta,\phi)=-e^{-i\pi\cos\theta(\sin\theta\cos\phi\sigma_x+\sin\theta\cos\phi\sigma_y+\cos\theta\sigma_z)}.
\end{equation}
Since the evolution operator $U^g$ already cancelled the dynamical phase, we do not need to additionally impose the condition of cancelling the dynamical phase. The solutions to maximizing the fidelity function for Not, Hadamard, Phase, and $\frac{\pi}{8}$ gate are listed below. Note that the solutions listed below are not unique.
\begin{itemize}
\item Not gate
\begin{align*}
&\begin{array}{c|cc}
\hline
& \mathrm{loop\ 1} & \mathrm{loop\ 2}\\
\hline
\omega/\Delta &1.591&1.755\\
\hline
\phi &2.253&4.180\\
\hline
\end{array}
\end{align*}

\item Hadamard gate
\begin{align*}
&\begin{array}{c|cc}
\hline
& \mathrm{loop\ 1} & \mathrm{loop\ 2}\\
\hline
\omega/\Delta &1.411&1.298\\
\hline
\phi &0.720&5.063\\
\hline
\end{array}
\end{align*}

\item Phase gate
\begin{align*}
&\begin{array}{c|cc}
\hline
& \mathrm{loop\ 1} & \mathrm{loop\ 2}\\
\hline
\omega/\Delta &1.492&1.492\\
\hline
\phi &3.725&2.940\\
\hline
\end{array}
\end{align*}

\item $\pi/8$ gate
\begin{align*}
&\begin{array}{c|cc}
\hline
& \mathrm{loop\ 1} & \mathrm{loop\ 2}\\
\hline
\omega/\Delta &1.398&1.398\\
\hline
\phi &3.695&3.302\\
\hline
\end{array}
\end{align*}
\end{itemize}
The gate length of the above four gates are $\frac{7.5268}{\Delta}$, $\frac{9.2908}{\Delta}$, $\frac{8.4213}{\Delta}$, and $\frac{8.9875}{\Delta}$, respectively. Apart from enlarging the Zeeman energy $\Delta$, we can also shorten the gate length by choosing solutions with larger $\omega$. For example, the phase gate can also be implemented by the following set of parameters, with the gate length $\frac{3.3448}{\Delta}$.
\begin{itemize}
\item Another set of parameters for Phase gate
\begin{align*}
&\begin{array}{c|cc}
\hline
& \mathrm{loop\ 1} & \mathrm{loop\ 2}\\
\hline
\omega/\Delta &3.757&3.757\\
\hline
\phi &2.921&5.277\\
\hline
\end{array}
\end{align*}
\end{itemize}
The gate length for phase gate is greatly shortened without varying the Zeeman energy. Hence, one can implement very fast HQGs by maximizing the fidelity function with large $\omega$. 

The two-qubit Hamiltonian is
\begin{align}\label{2bitHam}
\mathcal{H}_2&= \sum_{i=1}^2 \Omega_i (\mathrm{cos}(\omega_i t+\phi_i)\frac{\sigma_x^{i}}{2}+ \mathrm{sin}(\omega_i t+\phi_i)\frac{\sigma_y^{i}}{2} )  \\ \nonumber
& + \Delta_1 \frac{\sigma_z^{1}}{2} +\Delta_2 \frac{\sigma_z^{2}}{2}   + J\frac{\sigma_z^{1}\sigma_z^{2}}{4} ,
\end{align}
Values of parameters for the CNOT gate are listed below. The condition of cancelling the dynamical phase contains analytic solution to a quartic function, and hence we neglect the redundant closed-form expression here. For each pulse in the table, one can check that $\bra\phi_n(s)| H(s) |\phi_n(s)\ket=0$.
\begin{table}[!h]
\begin{tabular}{c|ccccccc}
\hline
Pulse & $\Omega_1/J$  & $\Omega_2/J$ & $\omega/J$ & $\phi_1$ & $\phi_2$ & $\Delta_1/J$ & $\Delta_2/J$ \\
\hline
P1 &1.446&4.131&8.478&3.111&1.590&0.268&4.168\\
\hline
P2 &1.956&3.819&7.837&4.437&1.431&0.561&3.761\\
\hline
P3 &3.394&4.339&8.745&2.053&3.467&1.836&3.702\\
\hline
P4 &1.807&3.591&7.394&5.127&4.532&0.510&3.555\\
\hline
P5 &2.551&4.015&8.183&1.172&4.864&0.967&3.797\\
\hline
\end{tabular}
\end{table}

\section{Proof of scalability}\label{appd:proof}
A general Hamiltonian that contains single-qubit RF pulse terms, single-qubit Zeeman terms and Ising-like coupling terms is
\begin{align}\notag
H=&\frac{1}{2}\sum_n (\Omega_n \cos(\omega_n t+\phi_n)\sigma^x_n+\Omega_n \sin(\omega_n t+\phi_n) \sigma^y_n))
\\
+&\sum_n \Delta_n \sigma^z_n+\frac{1}{4}\sum_{n<m}J_{nm}\sigma^z_n \sigma^z_m.
\end{align}
Following our assumption, the corresponding dynamical invariant can be written as
\begin{align}\notag
I=&\sum_n (\Omega_n \cos(\omega_n t+\phi_n)\sigma^x_n+\Omega_n \sin(\omega_n t+\phi_n) \sigma^y_n)
\\
+&\sum_n(\Delta_n-\omega_n) \sigma^z_i+\sum_{n<m}J_{nm}\sigma^z_n \sigma^z_m.
\end{align}
We can find that $I=2H-\sum_n\omega_n\sigma^z_n$. Therefore,
\begin{align*}
i [H,I ]&=i[H, 2H-\sum_n \omega_n \sigma^z_n]
\\
&=-i[H,\sum_n \omega_n \sigma^z_n]
\\
&=-i[\frac{1}{2}\sum_l \Omega_l \cos(\omega_l t+\phi_l)\sigma^x_l,\sum_n \omega_n \sigma^z_n]
\\
&\quad-i[\frac{1}{2}\sum_l \Omega_l \sin(\omega_l t+\phi_l) \sigma^y_l,\sum_n \omega_n \sigma^z_n]
\\
&=-\frac{i}{2}\sum_{ln}\Omega_l \omega_n \cos(\omega_l t+\phi_l)[\sigma^x_l,\sigma^z_n]
\\
&\quad-\frac{i}{2}\sum_{ln}\Omega_l \omega_n \sin(\omega_l t+\phi_l)[\sigma^y_l,\sigma^z_n]
\\
&=\sum_{ln}\Omega_l\omega_n(-\cos(\omega_l t+\phi_l)+\sin(\omega_l t+\phi_l))\delta_{ln}
\\
&=\sum_l \Omega_l\omega_l(-\cos(\omega_l t+\phi_l)+\sin(\omega_l t+\phi_l))
\\
&=-\frac{dI}{dt}.\qquad\qquad \mathrm{Q.E.D.}
\end{align*}


 \section{Evaluation of gate fidelity}
We perform quantum process tomography (QST) and Clifford-base randomized benchmarking (RB) for the non-adiabatic HQGs. In the following, we describe the process of implementing these two techniques in the NMR platform.

\emph{Quantum Process Tomography.} -- QPT is a conventional method to characterize the quality of a quantum channel. In NMR, we usually describe QPT in the Pauli basis. Assume that the quantum channel corresponding to the target gate is $\mathcal{U}$. For single-qubit channel $\mathcal{U}$, the map of $\mathcal{U}$ from the input state to the output state can be written as,
\be
\mathcal{U} \left(
          \begin{array}{c}
            I \\
            X \\
            Y \\
            Z \\
          \end{array}
        \right) = \left(
                    \begin{array}{cccc}
                      a^1_1 & a^1_2 & a^1_3 & a^1_4  \\
                     a^2_1 & a^2_2 & a^2_3 & a^2_4  \\
                    a^3_1 & a^3_2 & a^3_3 & a^3_4  \\
                    a^4_1 & a^4_2 & a^4_3 & a^4_4  \\
                    \end{array}
                  \right) \left(
          \begin{array}{c}
            I \\
            X \\
            Y \\
            Z \\
          \end{array}
        \right).
\ee
For a two-qubit channel $\mathcal{U}$,
\be
\mathcal{U} \left(
          \begin{array}{c}
            II \\
            IX \\
            ... \\
            ZZ \\
          \end{array}
        \right) = \left(
                    \begin{array}{ccccc}
                      a^1_1 & a^1_2& ... & a^1_{15}& a^1_{16} \\
                      a^2_1 & a^2_2& ... & a^2_{15}& a^2_{16} \\
                      ... & ... & ... & ... & ... \\
                      a^{16}_{1} & a^{16}_{2}& ... & a^{16}_{15}& a^{16}_{16} \\
                    \end{array}
                  \right) \left(
          \begin{array}{c}
            II \\
            IX \\
            ... \\
            ZZ\\
          \end{array}
        \right).
\ee
$\mathcal{U}$ is generally not a unitary channel due to the experimental errors. In the Pauli basis, the elements of $\mathcal{U}$ are real, and QPT needs to determine the unknown coefficients in $\mathcal{U}$ by measurement, such that $\mathcal{U}$ can be fully reconstructed. Taking the two-qubit channel as an example, we prepare the initial state to a Pauli basis, such as $IX$, and then apply $\mathcal{U}$  on it. The output is $\mathcal{U}(IX)=a^2_1II+a^2_2IX+...+a^2_{16}ZZ$. The coefficients $a^2_k$ $(k=1,2,..,16)$ can be measured by performing two-qubit quantum state tomography on the output state. The total numbers of experiments for reconstructing single-qubit and two-qubit channels, taking the normalization condition into account, are $4\times (4-1)=12$ and $16\times (16-1)=240$, respectively.

The sequence to prepare any Pauli state is as follows. At room temperature, the thermal equilibrium state of the NMR sample is a highly-mixed state, which is described by $\rho_{\rm thermal}=0.25II+\epsilon(4ZI+IZ)$ with the polarization $\epsilon$. Starting from this thermal state, we can easily prepare single-coherence terms with single-qubit rotations and gradients fields. Here, gradient field aims to crush the magnetization in the $xy$ plane. For instance, the term $ZI$ can be prepared by applying a $\pi/2$ rotation around the $y$ axis on $^{1}$H and a gradient field. The preparation of two-coherence terms needs the coupling evolution between $^{13}$C and $^{1}$H. For instance, the term $YZ$ can be prepared by the following sequence,
\begin{align}
YZ: ~~&\mathcal{R}^{2}_y(\pi/2) \rightarrow G \rightarrow \mathcal{R}^{1}_y(\pi/2) \rightarrow f(\frac{1}{2J}) .
\label{decompose}
\end{align}
where $\mathcal{R}^i_n(\theta)$ is a rotation about the axis $n$ with  angle $\theta$ acting on the $i$-th qubit and $f(1/2J)$ represents the free evolution of the system with the duration $1/2J$.

\emph{Clifford-based Randomized Benchmarking.} -- Compared with traditional QPT, RB can detect the error caused by the quantum gate after excluding the errors from the state preparation and measurement (SPAM). The workflow of RB contains two groups of experiments named by reference RB and interleaved RB sequences. The former is to measure the average error per Clifford gate and the latter aims to measure the error of a target gate. \\
(1) Reference RB. A random sequence including $m$ quantum gates $C'$s is firstly performed. $C$ is randomly chosen from the Clifford group. Then we deign a recovery gate $C_r$ to invert the system back to the input state and measure the survival probability of the input state. \\
(2) Interleaved RB. It is a similar sequence with the reference RB, where a target gate $G$ is interleaved after each Clifford gate $C$. A recovery gate $C'_r$ is designed to invert the whole sequence and the survival probability of the input state is also measured as the fidelity of this sequence. \\
The above sequence is randomly generated and is repeated for a certain number of times in experiments. The measured fidelities are averaged as $F$. $F$ is a function of the number of Clifford gates $m$ and it can be fitted by an exponential model $F(m)=Ap^m+B$. $p$ is the average decay of the sequence ($p=p_{\rm ref}$ for the reference RB and $p=p_{\rm gate}$ for the interleaved RB). $A$ and $B$ are the fitting coefficients  absorbing the SPAM errors.  Then we can calculate the fidelity of target gate by $F_{\rm gate}=1-(1-p_{\rm gate}/p_{\rm ref})(d-1)/d$, with $d=2^N$ for $N$ qubits. We implement the above RB for our single-qubit non-adiabatic HQGs including the X, H , P, and T gates. Random Clifford gates are chosen from a set of $I, R_x(\pm\pi/2), R_x(\pi),R_y(\pm\pi/2), $ and $R_y(\pi)$. A traceless Pauli operator $Z$ is chosen as the input state and we measure the the survival probability of $Z$ after the sequence.

\end{appendix}

\bibliography{Quantum}

\end{document}